%Paper: hep-ph/9212253
%From: "vaxrom::lusignoli"@tsmi19.sissa.it
%Date: Fri, 11 Dec 1992 12:23:25 +0100

%%%%%%%%%%%%%%%%%%%%%%%%%%%%%%%%%%%%%%%%%%%%%%%%%%%%%%%%%%%%%%%%%%%%%%%%%%%%
%     Two macros are needed: harvmac.tex and tables.tex                    %
%    They are available in the macro library, according to "LIST MACROS",  %
%      and therefore not included here.                                    %
%%%%%%%%%%%%%%%%%%%%%%%%%%%%%%%%%%%%%%%%%%%%%%%%%%%%%%%%%%%%%%%%%%%%%%%%%%%%

\input harvmac
\input tables
\def\KB{\overline{K}^0}
\def\PT{\widetilde P}
\def\singlespace{\baselineskip 12 pt}

\centerline{\titlerm {CP violating asymmetries in charged D meson decays}}
\vskip 2 cm
\centerline{F. Buccella$^{(a)}$, M. Lusignoli$^{(b)}$, G.Mangano$^{(a,c)}$,}
\vskip 5pt
\centerline{G. Miele$^{(a,c)}$, A. Pugliese$^{(b)}$ and P.
Santorelli$^{(a,c)}$}
\vskip 20pt
\centerline{
$^{(a)}$ {\it Dipartimento di Scienze Fisiche, Universit\`a di Napoli,
Napoli, Italy}}
\centerline{
$^{(b)}$ {\it Dipartimento di Fisica, Universit\`a ``La Sapienza'', Roma,}}
\centerline{{\it and I.N.F.N., Sezione di Roma I, Roma, Italy}}
\centerline{
$^{(c)}$ {\it I.N.F.N., Sezione di Napoli, Napoli, Italy}}
\vskip 6 truecm
\centerline {ABSTRACT}
\vskip 10truept
\singlespace\noindent
The CP violating asymmetries for Cabibbo suppressed charged D meson
decays in the standard model are estimated in the factorized
approximation, using the two-loop
effective hamiltonian and
a model for final state interactions previously tested for
Cabibbo allowed D decays. No new parameters are added. The predictions
are larger than expected and not too far from the experimental
possibilities.
\vfill\eject
\baselineskip 18 truept
\parindent=1cm

%\newsec{Introduction}
It is well known that CP violating effects show up in a decay process
only if the decay amplitude is the sum of two different
parts, whose phases are made of a weak (Cabibbo --
Kobayashi -- Maskawa) and a strong (final state interaction) contribution.
The weak contributions to the phases change sign when going to the
CP-conjugate process, while the strong ones do not.
Let us denote a generic decay amplitude of this type by
\eqn\eAMPL{{\cal A}= A\; e^{i\delta_1} + B\; e^{i\delta_2}}
and the corresponding
CP conjugate amplitude by
\eqn\eAMPB{{\bar{\cal A}}= A^* e^{i\delta_1} + B^* e^{i\delta_2}\,.}
The CP violating asymmetry in the decay rates will be therefore
\eqn\eCPV{a_{CP} \equiv
{|{\cal A}|^2-|{\bar{\cal A}}|^2 \over |{\cal A}|^2+|{\bar{\cal A}}|^2}
= {2\;\Im(A B^*) \, \sin(\delta_2-\delta_1) \over
|A|^2+|B|^2+2\;\Re(AB^*)\,\cos(\delta_2-\delta_1)}}
Both factors in the numerator of eq. \eCPV ~should be nonvanishing to have
a nonzero effect. Moreover, to have a sizeable asymmetry the moduli of the
two amplitudes $A$ and $B$ should not differ too much.

While the phases of the weak amplitudes are expressed in terms of
fundamental parameters of the theory (the angles and the CP violating phase
in the CKM quark mixing matrix), the strong phases $\delta_i$ are in
general unknown for heavy flavour decays.
For this reason the asymmetries in
charged $B$ (or $D$) decays are less studied and
less under control, theoretically
\foot{Although they are easier to measure
experimentally, due to their self-tagging property.}
, than
the asymmetries in mixing assisted channels in neutral
$B$ decays \ref\rBIGI{A. Carter and A.I. Sanda, Phys.Rev.Lett. 45 (1980) 952
\semi A. Carter and A.I. Sanda, Phys.Rev. D23 (1981) 1567 \semi
I.I. Bigi and A.I. Sanda, Nucl.Phys. B 193 (1981) 85.}.
In a previous paper \ref\rPRIMA{F.Buccella, M.Lusignoli, G.Miele and
A.Pugliese, Zeits.f.Phys. C 55 (1992) 243.}
we have presented a rather successful phenomenological analysis of
Cabibbo-allowed $D$ decays. In that analysis a specific model for final
state interactions \ref\rBUCC{F. Buccella, M. Forte, G. Miele and G. Ricciardi,
Zeits.f.Phys. C 48 (1990) 47.} was used, based on the assumption that
the scattering phase-shifts are dominated by the nearest resonance.

In this paper we exploit this model of final state interactions to
estimate the CP violating asymmetries for two-body $D$ decays in the
standard model.

These asymmetries have long be believed to be
unmeasurably small. For Cabibbo allowed decays, this is a
consequence of the smallness of
the second interfering amplitude with different weak phase ($B$ in
\eAMPL ), which can only be generated by mixing and doubly Cabibbo suppressed
decay. In fact, asymmetries in $D\to \overline{K}\pi$ have been recently
suggested
\ref\rOLIV{A. Le Yaouanc, L. Oliver and J.-C. Raynal, preprint LPTHE 92/34,
Phys.Lett. B (to be published).}
as a sensitive tool to observe possible ``new physics'' effects in
CP violating amplitudes, taking advantage of the large strong phase
differences.

In the Cabibbo suppressed case the penguin diagram
\ref\rSVZ{M. Shifman, A. Vainshtein and Zakharov, Sov.Phys. JETP
45 (1977) 670.} contribution to the effective Hamiltonian,
although its coefficient is rather small,
provides a nonnegligible
second amplitude. We will
therefore limit our calculations to Cabibbo suppressed decays.
Moreover, only {\it charged} $D$ will be considered. In fact,
to describe final state
interactions in $D^0$ decays, that receive contributions
from not yet observed isoscalar resonances,
new undetermined parameters would be needed.

As we will see, the asymmetries are small but larger
than previously expected
\ref\rCHAU{L.-L. Chau and H.-Y. Cheng, Phys.Rev.Lett. 53 (1984) 1037.},
and in some cases they are O($few\; 10^{-3}$).
To observe at $3\,\sigma$ level an asymmetry of $3 \cdot 10^{-3}$
in a channel having a decay branching ratio of 0.5\%,
a number of charged $D$ mesons equal to $2 \cdot 10^8$ is required,
{\it with 100\% efficiencies}.
Present projects of tau-charm factories do not
have a luminosity high enough to allow such a
measurement, but it is not unconceivable that one could
reach this sensitivity in a not too far future.

\vskip 0.5 cm
%\newsec{Cabibbo suppressed decay amplitudes in the factorized approximation}
The effective weak hamiltonian for Cabibbo suppressed nonleptonic decays of
charmed particles is given by
\eqn\eHEFF{\eqalign{ H_{eff} &=\;
       {G_F \over \sqrt 2}\,V_{ud}\,V_{cd}^*\;
      \bigl[C_1\;Q_1^d\;+\;C_2\;Q_2^d \bigr]\; +\cr
    &+\;   {G_F \over \sqrt 2}\,V_{us}\,V_{cs}^*\;
      \bigl[C_1\;Q_1^s\;+\;C_2\;Q_2^s \bigr]\; -\cr
    &-\;   {G_F \over \sqrt 2}\,V_{ub}\,V_{cb}^*
      \; \sum_{i=3}^6 \;C_i\; Q_i \;\;\;+\;\; {\rm h.c.}\,,\cr}}
In eq. \eHEFF ~the operators are \ref\rGILMAN{F.J. Gilman and M.B. Wise,
Phys.Rev. D 20 (1979) 2392.}:
\eqn\eOPER{\eqalign{
Q_1^d =&\;
       \bar{u}^\alpha\,\gamma_\mu\,(1-\gamma_5)\,d_\beta\;
       \bar{d}^\beta\,\gamma^\mu\,(1-\gamma_5)\,c_\alpha, \cr
Q_2^d =&\;
       \bar{u}^\alpha\,\gamma_\mu\,(1-\gamma_5)\,d_\alpha\;
       \bar{d}^\beta\,\gamma^\mu\,(1-\gamma_5)\,c_\beta, \cr
Q_3 =&\;
       \bar{u}^\alpha\,\gamma_\mu\,(1-\gamma_5)\,c_\alpha\;
      \sum_q\; \bar{q}^\beta\,\gamma^\mu\,(1-\gamma_5)\,q_\beta, \cr
Q_4 =&\;
       \bar{u}^\alpha\,\gamma_\mu\,(1-\gamma_5)\,c_\beta \;
       \sum_q\;\bar{q}^\beta\,\gamma^\mu\,(1-\gamma_5)\,q_\alpha, \cr
Q_5 =& \;
       \bar{u}^\alpha\,\gamma_\mu\,(1-\gamma_5)\,c_\alpha\;
       \sum_q\;\bar{q}^\beta\,\gamma^\mu\,(1+\gamma_5)\,q_\beta, \cr
Q_6 =& \;
       \bar{u}^\alpha\,\gamma_\mu\,(1-\gamma_5)\,c_\beta \;
       \sum_q\;\bar{q}^\beta\,\gamma^\mu\,(1+\gamma_5)\,q_\alpha\;. \cr}}

The operator $Q_1^s$ ($Q_2^s$) in eq. \eHEFF ~is obtained from
$Q_1^d$ ($Q_2^d$) with
the substitution $(d \to s)$. In eq.
\eOPER ~$\alpha$ and $\beta$ are colour indices and in the ``penguin''
operators
$q$ ($\bar{q}$) is to be summed over all active flavours ($u$, $d$, $s$).

We have evaluated the coefficients $C_i$ using the two-loop
anomalous dimension matrices recently calculated by A. Buras and
collaborators
\ref\rBURAS{A.J. Buras, M. Jamin, M.E. Lautenbacher and P.E. Weisz,
 Nucl.Phys. B 370 (1992) 69; Nucl.Phys. B 375 (1992) 501 (addendum).}.
The effect of next-to-leading corrections is particularly large for the
coefficients of the penguin operators, as it has already been noted in
ref. \rBURAS ~in the case of $\Delta S=1$ decays. One problem that we have
to face is the renormalization scheme dependence of the coefficients,
that should be canceled by a corresponding dependence in the matrix
elements of the operators. Unfortunately, we estimate the matrix elements
of four-fermion operators using a specific model, and it is not evident
to which scheme this corresponds
\foot{ An analogous problem for the dependence of the coefficients
on the subtraction point, $\mu$, already appears at leading order.}.
We calculated the coefficients in both schemes (naive dimensional
regularization and 't Hooft - Veltman scheme) for which anomalous dimension
matrices have been given. We follow then the prescription of
Buras et al., given in eq. (3.6) of ref. \rBURAS , to define ``scheme
independent'' coefficients. Although not unique, this prescription is
phenomenologically favoured, in that it reinforces the results of the
leading order calculation and provides large penguin contributions, that
are welcome in the $\Delta S=1$ case. It turns out that this prescription
gives for the dominant penguin operator coefficient $C_6$ a value that is
about two and a half times larger than the leading order prediction
$C_6^{LL}$. We stress
that the results for $C_6$ in both schemes are also considerably larger
than $C_6^{LL}$.

In ref. \rPRIMA ~we fitted the value of $\Lambda_4^{\overline {MS}}$,
among other parameters, and
obtained the result
$\Lambda_4^{\overline {MS}} \sim 200\,{\rm MeV}$. We therefore
adopted this value in the calculation of the coefficients $C_i$.
The results are reported in Table 1.

Given the effective Hamiltonian, to obtain the decay amplitude
\eqn\eAAAA{ {\cal A}_{\rm w}\;(D \to f) \;=\; <f| H_{eff} |D>}
we evaluate the matrix elements of the
operators  in the factorization approximation, neglecting moreover
the colour-suppressed contributions, following
\ref\rBSW{M. Bauer and B. Stech,
Phys.Lett. 152B (1985) 380 \semi M. Bauer, B. Stech and M. Wirbel,
Zeits.f.Phys. C34 (1987) 103.}
and \rPRIMA . We also neglect the contribution
of the dimension five operator
\eqn\eMOMMAG{\overline{Q}=\;m_c
       \bar{u}^\alpha\,\sigma^{\mu \nu}\,(1+\gamma_5)\;t^A_{\alpha \beta}
\,c^\beta\;G^A_{ \mu \nu}\;,}
that is generated at the two-loop level. Its
Wilson coefficient has not been evaluated in \rBURAS , where
the masses of external quarks have been neglected. Its matrix element
cannot be calculated in the factorized approximation and it would be
anyhow suppressed by $1/N_c$ in the large $N_c$ limit.

As an illustration, we discuss the explicit example
$D^+\to K^+ \overline{K}^{*0}$-- a decay channel that turns out to have a
sizeable asymmetry.
The expressions for the matrix
elements of the operators $Q_i$ ($i$=1,\dots 5) are easily
obtained following ref. \rPRIMA ~and are:
\eqn\eEXAMP{\eqalign{
<K^+ \overline{K}^{*0}|Q_1^d|D^+> \;=&
<K^+ \overline{K}^{*0}|Q_1^s|D^+> = 0\;,\cr
<K^+ \overline{K}^{*0}|Q_2^d|D^+> \;=&
- 2\, M_{K^*}\;(\epsilon^* \cdot p_K)\;(m_u+m_d)\;W_{PV}\;, \cr
<K^+ \overline{K}^{*0}|Q_2^s|D^+> \;=&
\;\, 2\, M_{K^*}\;(\epsilon^* \cdot p_K)\;
f_K\;{a_{cs} \over 1\,-\,M_K^2/M^2_{D_{s}}}\;, \cr
<K^+ \overline{K}^{*0}|Q_4|D^+> \;=&
<K^+ \overline{K}^{*0}|Q_2^d|D^+>\,+\,
<K^+ \overline{K}^{*0}|Q_2^s|D^+>\;,\cr
<K^+ \overline{K}^{*0}|Q_3|D^+> =&
<K^+ \overline{K}^{*0}|Q_5|D^+> = 0\;.\cr}}
In eq. \eEXAMP , in addition to self-explanatory symbols for particle masses
and momenta and for the $K^*$ polarization vector $\epsilon$, two quantities
need a definition: $a_{cs}\simeq0.8$ is the axial charge
and $W_{PV}$ parameterizes the annihilation contribution.
$W_{PV}$ has been determined by a fit to the
Cabibbo allowed decay rates in ref. \rPRIMA ,
with the result $W_{PV}= 0.53 \pm 0.13$
\foot{ We will
assume the central values for the parameters determined in ref. \rPRIMA ~in
the following estimates.}.
The operator $Q_6$ must first be
written in a Fierz-rearranged form:
\eqn\eQSEI{Q_6 = \;- 2 \;\bar{q}^\beta (1-\gamma_5) c_\beta \bar{u}^\alpha
(1+\gamma_5) q_\alpha .}
In the factorization approximation, its matrix elements are given in
terms of matrix elements of scalar and pseudoscalar densities. These
in turn may be related to the matrix elements of the divergences of
vector and axial currents, and thereby to form factors that have been
discussed in \rPRIMA . The result is:
\eqn\eEXUMP{\eqalign{<K^+ \overline{K}^{*0}|Q_6|D^+> \;=
\; 4 M_{K^*}\;(\epsilon^* \cdot p_K) &\biggl[
{M_D^2 \over (m_c+m_d)}\;W_{PV}\;- \cr
&-\,{M_K^2\;f_K \over
(m_s+m_u)(m_c+m_s)} \cdot
{ a_{cs} \over 1\,-\,M_K^2/M^2_{D_{s}}}
   \biggr].\cr}}

We note that the penguin operator $Q_6$ has, by far,
the largest matrix element.
Even if the coefficient $C_6$ is considerably smaller than $C_2$ or $\;C_1$
its contribution would be dominant, were it not for the smallness of the
CKM factor: $|V_{ub}V_{cb}^*|/|V_{us}V_{cs}^*|\simeq 10^{-3}$.
In fact, the $Q_6$ penguin operator gives the dominant contribution
to the {\it imaginary part} of the amplitude ${\cal A}_{\rm w}$, since all
the CKM factors' imaginary parts are of the same order.

Expressions analogous to \eEXAMP ~and \eEXUMP ~
can be derived for the other two-body Cabibbo first
forbidden decay channels of $D^+$ and $D^+_s$.

At this stage one would not have any
CP violating asymmetry, since no strong
phases from final state interactions have been included
in the amplitudes ${\cal A}_{\rm w}\;(D \to f)$.
Rescattering, to be discussed in the next
section, will provide the necessary nonzero strong phase-shifts.

\vskip 0.5 cm
%\newsec{Strong phase shifts and CP violating asymmetries}
We make the assumption that the final state
interactions are dominated by resonant contributions
\rBUCC .

In the mass region of pseudoscalar charmed particles there is evidence,
albeit not very strong
\ref\rPDG{K. Hikasa {\it et al.} (Particle Data Group), Phys.Rev.
D 45 (1992) Part II.} , for a $J^P=0^+$ resonance $K^*_0$ (with mass
1950 MeV, width 201$\pm$86 MeV and 52\% branching ratio in $K \pi$
\ref\rASTON{D. Aston et al., Nucl.Phys. B296 (1988) 493.}),
a $J^P=0^-$ $K(1830)$ (with $\Gamma = 250$ MeV and
an observed decay to $K \phi$
\ref\rARMST{T. Armstrong et al., Nucl.Phys. B221 91983) 1.})
and a $J^P=0^-$ $\pi(1770)$ with $\Gamma = 310$ MeV
\ref\rBELL{G. Bellini et al., Phys.Rev.Lett. 48 (1982) 1697.}.
These resonances should dominate, respectively, the rescattering
effects for Cabibbo
forbidden decays of $D_s^+\to PP$, $D_s^+\to PV$ and $D^+\to PV$, where
$P(V)$ indicates a pseudoscalar (vector) decay product.
In \rPRIMA ~we assumed the existence of a $J^P=0^+$ resonance $a_0$,
that should be relevant for Cabibbo forbidden $D^+\to PP$ decays,
with mass 1890 MeV and width $\sim 200$ MeV.
We will not try to
discuss the Cabibbo forbidden decay amplitudes for $D^0$ mesons;
in this case several other unobserved isoscalar nonstrange
resonances should determine the final state interactions and this
would bring new unknown parameters in the calculations.

The FSI effect modifies the amplitudes for Cabibbo suppressed $D^+$
and $D_s^+$ decays
in the following way \rBUCC :
\eqn\eRESCA{\eqalign{{\cal A} (D \to V_h\,P_k) &= {\cal A}_{\rm w}
  (D \to V_h\,P_k) + c_{hk}[\exp(i\delta_8)-1]
   \sum_{h\prime k\prime} c_{h\prime k\prime}\,{\cal A}_{\rm w}
  (D \to V_{h\prime}\,P_{k\prime})= \cr
&= d_{hk}\,{\cal A}^{27} + c_{hk}\,\exp(i\delta_8)\,{\cal A}^8\;.}}
In \eRESCA ~ $c_{hk}$ [$d_{hk}$] are the normalized ($\sum c_{hk}^2 = 1$)
couplings of a pseudoscalar ($P$) and a vector ($V$) octets to
a pseudoscalar octet [27-plet]. The resonance $\PT$ is present only in
the octet case, and one has
\eqn\ePHSH{\sin \delta_8 \, \exp (i \delta_8) =
   {\Gamma ({\PT} ) \over 2\,(m_{\PT} - m_D) - i\,\Gamma ({\PT} )}.}
Similar expressions hold for ${\cal A} (D \to P_h\,P_k)$.

The CP violating asymmetry of eq. \eCPV ~is given by:
\eqn\eASYMM{a_{CP} = {2\;\sin \delta_8 \; c_{hk}\,d_{hk} \;
\Im({\cal A}^{27}\;{\cal A}^{8*})
\over c_{hk}^2\,|{\cal A}^8|^2 + d_{hk}^2\,|{\cal A}^{27}|^2 +
2\; c_{hk}\,d_{hk}\;\Re({\cal A}^{27}\;{\cal A}^{8*})
\cos \delta_8}}
To have a nonzero asymmetry the weak phases of ${\cal A}^8$ and ${\cal A}^{27}$
must be different. The phase difference is provided essentially by the
penguin contributions, which dominate the imaginary parts of the amplitudes,
and appear in ${\cal A}^8$ only.
On the other hand, it is true that in some channels there
is no rescattering in our model and $a_{CP}=0$. This is the case for instance
for the decay amplitude ${\cal A}(D^+ \to \pi^+ \pi^0)$, which only
contains the term ${\cal A}^{27}$.

A particular treatment is needed for those decay channels where  neutral
kaons appear. In fact, these particles will be mostly detected through
their decays into two pions, and these decays are by
themselves affected by CP violation (in the $K$ system). One has
therefore to be able to disentangle the CP violating effects in $D$
and $K$ decays. Define the ``experimental'', time-integrated asymmetry
for the $D^+$ decay to one charged and one neutral $K$ meson:
\eqn\eASTI{\Delta(T) = {\int_0^T [\Gamma(D^+ \to K^+\overline{K}^0\to
K^+\pi^+\pi^-)(t) - \Gamma(D^-\to K^-K^0 \to
K^-\pi^+\pi^-)(t)] dt \over \int_0^T [\Gamma(D^+\to K^+\overline{K}^0\to
K^+\pi^+\pi^-)(t) + \Gamma(D^-\to K^-K^0\to
K^-\pi^+\pi^-)(t)] dt}}
and recall the usual definitions of parameters for the neutral kaon decays
\eqn\eKCP{\eqalign{\Delta M =& M_{K_L}- M_{K_S} \;,\cr
\Gamma =& {1\over2}(\Gamma_S + \Gamma_L)\;, \cr
\eta_{+-} =& {A(K_L\to \pi^+\pi^-) \over A(K_S\to \pi^+\pi^-)}=
\epsilon + \epsilon'\;.\cr}}
It is also convenient to define
\ref\rAMELI{G. Amelino-Camelia, F. Buccella, G. D'Ambrosio, A. Gallo,
G. Mangano and M.Miragliuolo, Zeits.f.Phys. C 55 (1992) 63.}:
\eqn\eNOTAZ{\eqalign{
f(T)\;=&\;{2 \over (\Delta M)^2+ \Gamma^2} \biggl[
(\Gamma \;\Re\,\eta_{+-}\;+\;\Delta M \;\Im\,\eta_{+-})\;
(1-e^{-\Gamma T}\;\cos \Delta M\,T)\;+ \cr
&\phantom{\;{2 \over (\Delta M)^2+ \Gamma^2} \biggl[}
+\;(\Delta M \;\Re\,\eta_{+-}\;-\;\Gamma \;\Im\,\eta_{+-})
\;e^{-\Gamma T}\;\sin \Delta M T\; \biggr]\;, \cr
g(T)\;=&\;{\Gamma_L \; \Gamma_S \;f(T) \over \Gamma_L\;(1-e^{-\Gamma_S T})
+ \Gamma_S \;(1-e^{-\Gamma_L T})\;|\eta_{+-}|^2}\;, \cr
\gamma\;\;\simeq &\;
a_{CP}+\;2\;\Re\,\epsilon\;. \cr}}

The resulting asymmetry is:
\eqn\eASYMTI{\Delta(T) = {\gamma - g(T) \over 1\;-\;\gamma\;g(T)}}
It is instructive to consider the limits of this expression for short
($<<\tau_S$) and long ($>>\tau_S$) time $T$:
\eqn\eLIMIT{\eqalign{
\displaystyle\lim_{T\to 0}\Delta(T)\;=&\;a_{CP} \cr
\displaystyle\lim_{T\to \infty}\Delta(T)\;\simeq&\;a_{CP}\;-
\;2\;\Re\epsilon \;.\cr}}
Eq. \eLIMIT ~shows that to perform a measurement of the $D^+$ decay asymmetry
in these channels will be essentially hopeless if the asymmetry
$a_{CP}$ turns out to be
considerably less than $2\;\Re\,\epsilon$, the asymmetry in $K_L$ semileptonic
decays.

For $D_s$ decays, the equation analogous to \eASTI ~would be
\eqn\eASTS{\Delta_s(T) = {\int_0^T [\Gamma(D_s^+ \to \pi^+K^0\to
\pi^+\pi^+\pi^-)(t) - \Gamma(D_s^-\to \pi^-\overline{K}^0 \to
\pi^-\pi^+\pi^-)(t)] dt \over \int_0^T [\Gamma(D_s^+ \to \pi^+K^0\to
\pi^+\pi^+\pi^-)(t) + \Gamma(D_s^-\to \pi^-\overline{K}^0 \to
\pi^-\pi^+\pi^-)(t)] dt}\;.}
The resulting expression can be obtained from \eNOTAZ ~and
\eASYMTI ~by changing the signs of the terms containing $\epsilon$
and $\eta_{+-}$. Therefore
\eqn\eLIMIS{\displaystyle\lim_{T\to \infty}\Delta_s(T)\;\simeq \;a_{CP}\;+
\;2\;\Re\epsilon \;.}

\vskip 0.5 cm
%\newsec{Results and discussion}
The decay amplitudes for Cabibbo-suppressed charm decays (and especially
their CP violating asymmetries) depend on all the entries in the first
two rows of the quark mixing matrix. In the Wolfenstein parametrization
\ref\rWOLFEN{L. Wolfenstein, Phys.Rev.Letters 51 (1983) 1945.}
, extended to keep terms up to $O(\lambda^5)$ in the imaginary parts,
the matrix is written as follows:
\eqn\eCKMMATR{V\;=\;
\pmatrix{1-{1\over 2}\lambda^2 & \lambda & A\,\lambda^3\,\rho\,e^{-i\delta} \cr
-\lambda\,(1+A^2\,\lambda^4\,\rho\,e^{i\delta}) & 1-{1\over 2}\lambda^2 &
A\,\lambda^2 \cr
A\,\lambda^3\,[1-(1-{1\over 2}\lambda^2)\,\rho\,e^{i\delta}] &
-A\,\lambda^2\,(1+\lambda^2\,\rho\,e^{i\delta}) & 1 \cr}}
In \eCKMMATR , $\lambda = 0.220$ is the Cabibbo angle, we have
chosen $V_{cb}\;=\;A\;\lambda^2 = 0.046$ and allowed
different values for $\rho$, $\rho=0.50\pm 0.14$, and for the
CP violating phase $\delta$. Following the analysis of
\ref\rLMMR{M. Lusignoli, L. Maiani, G. Martinelli and L. Reina,
Nucl.Phys. B 369 (1992) 139.}
and assuming for the top quark mass the ``favourite'' value (140 GeV),
we considered both positive ($0.47\div 0.81$) and negative
($-0.98 \div -0.80$) values for $\cos \delta$, corresponding to the
two possible shapes of the unitarity triangle compatible with present
data on $B-\bar{B}$ mixing and the value of the $\epsilon$ parameter.
The positive $\cos \delta$ values correspond to
larger CP violating effects in $K$ decays ($\epsilon'/\epsilon$
and charged $K$ decay asymmetries),
in $B$ decays and also in $D$ decays.

We have evaluated the branching ratios for decay into twelve different
two-particle final states, and the corresponding CP violating asymmetries,
both for $D^+$ and for $D^+_s$.
The results are reported in Tables 2 and 3, respectively.

We note that the predicted
branching ratios are in reasonable agreement with the existing data
\rPDG , listed in the second columns, which
have {\it not} been used in fitting the parameters.
The variations in the theoretical predictions with $\rho$ and
$\cos \delta$ are less than $10^{-4}$ and therefore neglected in
column three.

We also note that very recently a doubly Cabibbo-suppressed decay rate
has been measured for the first time
\ref\rNEW{J.C. Anjos {\it et al.}, Phys.Rev.Lett. 69 (1992) 2892},
with the result
\eqn\eDCABF{{\Gamma (D^+ \to \phi K^+) \over \Gamma (D^+ \to \phi \pi^+)}
\vert _{exp.} = \bigl(5.8^{\displaystyle +3.2}_{\displaystyle -2.6} \pm
0.7 \bigr) \cdot 10^{-2}\;.}
With the parameters determined in \rPRIMA ~our model predicts
\eqn\eDCFT{{\Gamma (D^+ \to \phi K^+) \over \Gamma (D^+ \to \phi \pi^+)}
\vert _{th.} = 3.2 \cdot 10^{-2}\;.}
This ratio agrees within one standard deviation with the experimental result
\eDCABF , however it should be noted that the predicted individual rates
are smaller than the data by a somewhat larger amount, see Table 2.

The errors in the asymmetries reported in columns four and five
have been estimated by semi-dispersion in the numbers obtained
varying one at a time $\rho$ and
$\cos \delta$ in the ranges previously mentioned. The asymmetries
for an obtuse $\delta$ (column five) are generally smaller by
a factor two. There are some reasons to prefer the acute-angled
solution for the unitarity triangle, though. Lattice QCD calculations
\ref\rLATTICE{C.R. Allton {\it et al.}, Nucl.Phys. B 349 (1991) 598 \semi
C.Alexandrou {\it et al.}, Phys. Lett. B 256 (1991) 60 \semi
C.R.Allton {\it et al.}, Nucl.Phys. B (Proc. Suppl.) 20 (1991) 504 \semi
A.Abada {\it et al.}, Nucl.Phys. B 376 (1992) 172 \semi
C.Bernard, C.Heard, J.Labrenz and A.Soni, Nucl.Phys.
B (Proc.Suppl.) 26 (1992) 384.},
as well as recent evaluations with QCD sum rules
\ref\rNARISON{S. Narison, QCD Spectral Sum Rules, World Scientific
Lecture Notes in Physics, 26 (World Scientific, Singapore, 1989) \semi
M.Neubert, Phys.Rev. D 45 (1992) 2451.},
favour a rather large value for the $B$ meson decay constant $f_B$.
This, together with the large top quark mass indicated
by the electroweak data analyses
\ref\rMTOP{J. Ellis and G.L. Fogli, Phys.Lett. B 249 (1990) 543 \semi
P. Langacker and M.-X. Luo, Phys.Rev. D 44 (1991) 1591 \semi
A. Borrelli, L. Maiani and R. Sisto, Phys.Lett. B 244 (1990) 117.},
would select an acute-angled unitarity triangle.

Hoping that nature is kind to us and $\cos \delta > 0$,
our results predict
the largest asymmetries ($\sim 10^{-2}$) for some $D_s$
rare decay channels.
Considering however that $D_s$ are also difficult to produce,
we consider the
$D^+$ decay channels having the
highest asymmetries ($\sim 3\cdot 10^{-3}$) as more promising
to observe CP violating effects.

The best candidate is probably the decay $D^+ \to K^+ \overline{K}^{*0}$.
Here the final state contains three charged particles 2/3 of the times,
and the decay branching ratio
has been already measured to be half a percent.
These are the numbers we used in the discussion at the end of
the introduction. A tau-charm factory at the $\psi''(3770)$
with an integrated luminosity ${\cal L} = 10^{40}\;{\rm cm}^{-2}$
would produce about $10^8$ charged $D$ mesons and therefore $\sim 1.5\cdot
10^5$ decay events $D^+ \to K^+ \overline{K}^{*0} \to K^+ K^- \pi^+$
and as many $D^- \to K^- K^{*0} \to K^- K^+ \pi^-$. Without including
the further reduction due to actual experimental efficiencies, this
number of events would imply for the asymmetry a statistical accuracy of
$1.7 \cdot 10^{-3}$. As we already said, the assumed luminosity
is not enough to establish the effect, but at the same time not too far
from this goal.

\vfill\eject

\vskip 4 cm
\begintable
  | leading order | naive dim. reg. | 't Hooft-Veltman | ``scheme indep.'' \cr
$C_1$ | -0.439  | -0.303  | -0.311  | -0.514  \cr
$C_2$ | 1.216   | 1.139   | 1.032   | 1.270   \cr
$C_3$ | 0.0073  | 0.0155  | 0.0126  | 0.0202  \cr
$C_4$ | -0.0185 | -0.0398 | -0.0291 | -0.0460 \cr
$C_5$ | 0.0057  | 0.0099  | 0.0084  | 0.0127  \cr
$C_6$ | -0.0210 | -0.0447 | -0.0320 | -0.0541
\endtable
\vskip 1 cm
\centerline{TABLE 1}
\vskip 0.5 cm
\noindent
Coefficients in the effective Hamiltonian, eq. \eHEFF , for
$\Lambda_4^{\overline {MS}} = 200 \;{\rm MeV}$ and $m_b = 4.8\;{\rm GeV}$,
at a scale
$\mu = m_c = 1.5\;{\rm GeV}$, in different renormalization schemes.
See text for further details.
\vfill\eject
\vskip 4 cm
\begintable
decay channel |
$^{\displaystyle 10^2 \times B.R.}_{\displaystyle \;\;\;(exp.)}$ |
$^{\displaystyle 10^2 \times B.R.}_{\displaystyle \;\;\;\;(th.)}$ |
$^{\displaystyle 10^3 \times a_{CP}}_{\displaystyle \cos \delta > 0}$ |
$^{\displaystyle 10^3 \times a_{CP}}_{\displaystyle \cos \delta < 0}$ \cr
$D^+ \to \pi^+ \pi^0$ | $<.53$ | .21 | --- | --- \cr
$D^+ \to \pi^+ \eta$  | $.66 \pm .22$| .78 | $-1.5\pm 0.4$ | $-0.7\pm 0.4$ \cr
$D^+ \to \pi^+ \eta'$ | $<.8$  | .10 | $0.04\pm 0.01$ | $0.02\pm 0.01$ \cr
$D^+ \to K^+ \KB$ | $.73 \pm .18$ | 1.11 | $1.0 \pm 0.3$ | $0.5 \pm 0.3$ \cr
$D^+ \to \rho^0 \pi^+$ | $<.12$ | .20 | $-2.3 \pm 0.6$ | $-1.2 \pm 0.6$ \cr
$D^+ \to \rho^+ \pi^0$ |    | .39 | $2.9 \pm 0.8$ | $1.5 \pm 0.8$ \cr
$D^+ \to \rho^+ \eta$  | $<1.0$| .12 | --- | ---  \cr
$D^+ \to \rho^+ \eta'$ | $<1.4$ | .07 | --- | ---  \cr
$D^+ \to \omega \pi^+$ | $<.6$ | .07 | --- | ---  \cr
$D^+ \to \phi \pi^+$ | $.60 \pm .08$ | .36 | --- | ---  \cr
$D^+ \to K^{*+} \KB$ |    | 1.85 | $-0.9 \pm 0.3$ | $-0.5 \pm 0.3$  \cr
$D^+ \to \overline{K}^{*0} K^+$ | $.47 \pm .09$ | .36 | $2.8 \pm 0.8$ |
$1.4 \pm 0.7$
\endtable
\vskip 1 cm
\centerline{TABLE 2}
\vskip 0.5 cm
\noindent
Branching Ratios and CP-violating decay asymmetries for $D^+$ Cabibbo
forbidden decays. Experimental data and 90\% c.l. upper bounds taken
from ref. \rPDG .
\vfill\eject
\vskip 4 cm
\begintable
decay channel |
$^{\displaystyle 10^2 \times B.R.}_{\displaystyle \;\;\;(exp.)}$ |
$^{\displaystyle 10^2 \times B.R.}_{\displaystyle \;\;\;\;(th.)}$ |
$^{\displaystyle 10^3 \times a_{CP}}_{\displaystyle \cos \delta > 0}$ |
$^{\displaystyle 10^3 \times a_{CP}}_{\displaystyle \cos \delta < 0}$ \cr
$D^+_s \to K^+ \pi^0$ |   | .12 | $-0.5 \pm 0.2$ | $-0.2 \pm 0.1$ \cr
$D^+_s \to K^+ \eta$  |   | .004 | $2.3 \pm 0.7$ | $1.2 \pm 0.6$ \cr
$D^+_s \to K^+ \eta'$ |   | .80 | $1.0 \pm 0.3$ | $0.5 \pm 0.3$ \cr
$D^+_s \to K^0 \pi^+$ | $<.6\,^{(a)}$ | .48 | $-1.5 \pm 0.5$ |
 $-0.8 \pm 0.4$ \cr
$D^+_s \to \rho^+ K^0$ |   | 1.59 | $0.6 \pm 0.2$ | $0.3 \pm 0.2$ \cr
$D^+_s \to \rho^0 K^+$ |   | .15 | $1.0 \pm 0.3$ | $0.5 \pm 0.3$ \cr
$D^+_s \to \omega K^+$ |   | .03 | $-5.5 \pm 1.6$ | $-2.7 \pm 1.4$ \cr
$D^+_s \to \phi K^+$ | $<.2\,^{(b)}$| .18 | $0.14 \pm 0.04$ |
 $0.07 \pm 0.04$ \cr
$D^+_s \to K^{*+} \pi^0$ |   | .03 | $-5.8 \pm 1.7$ | $-2.9 \pm 1.5$ \cr
$D^+_s \to K^{*+} \eta$  |   | .05 | $-3.5 \pm 1.0$ | $-1.7 \pm 0.9$ \cr
$D^+_s \to K^{*+} \eta'$ |   | .01 | $-8.1 \pm 2.3$ | $-4.1 \pm 2.1$ \cr
$D^+_s \to K^{*0} \pi^+$ |   | .23 | $-2.6 \pm 0.8$ | $-1.3 \pm 0.7$
\endtable
\vskip 1 cm
\centerline{TABLE 3}
\vskip 0.5 cm
\noindent
Branching Ratios and CP-violating decay asymmetries for $D^+_s$ Cabibbo
forbidden decays. Experimental 90\% c.l. upper bounds are taken:
($a$) from ref. \rPDG ~and ($b$) from ref. \rNEW .
\vfill\eject

\listrefs

\bye